\documentclass[11pt,twoside]{article}

\usepackage{asp2006}
\usepackage{epsf}
\usepackage{graphicx}
\usepackage{lscape}

\begin{document}
\title{
  The Evolution of the Visible and Hidden Star Formation in the Universe:
  Implication from the Luminosity Functions at FUV and FIR
}
\author{T. T. Takeuchi\altaffilmark{1}, V. Buat\altaffilmark{2}, 
  and D. Burgarella\altaffilmark{2}
}
\altaffiltext{1}{Astronomical Institute, Tohoku University, Sendai, Japan}
\altaffiltext{2}{Laboratoire d'Astrophysique de Marseille, Marseille, France}

\begin{abstract}
Based on {\sl GALEX} and {\sl IRAS}/{\sl Spitzer} datasets,
we have found that both FUV and FIR luminosity functions (LFs) 
show a strong evolution from $z=0$ to $z=1$, but the FIR LF
evolves much stronger than the FUV one.
Consequently, the FIR/FUV luminosity density ratio increases from 
4 ($z=0$) to 15 ($z=1$).
It means that more than $80\;\%$ of the star-forming activity 
in the Universe is hidden by dust at $z=1$.
{}To explore this issue further, we have performed a combined analysis of 
the galaxy sample in FUV and FIR. 
For the Local Universe we used {\sl GALEX}-{\sl IRAS} sample,
whereas at $z=1$ we used the Lyman-break galaxy sample selected
by {\sl GALEX} bands constructed by Burgarella et al.\ (2005),
which is known to be representative of visible (i.e., non-obscured) 
star-forming galaxies at $z=1$.
{}From these datasets, we constructed the LFs of the FUV-selected 
galaxies by the survival analysis to, take into account the 
upper-limit data properly.
We discovered that the FIR LF of the Lyman-break galaxies show 
a significant evolution comparing with the local FIR LF,
but it is a factor of $2\mbox{--}3$ lower than the global FIR LF
\citep{lefloch05}.
This indicates that the evolution of visible galaxies is not
strong enough to explain the drastic evolution of the FIR LF.
Namely, a FIR-luminous, rapidly diminishing population of galaxies is 
required.
\end{abstract}

\vspace{-0.5cm}
\section{Introduction}

Newly formed massive stars emit strong far-ultraviolet (FUV) radiation.
However the attenuation of the FUV light by the interstellar dust is 
a major issue to derive quantitative SFR from the FUV, even at low-$z$
\citep[e.g.,][]{buat05}
On a global point of view, the recent observations conducted by {\sl Spitzer} 
and {\sl GALEX} have allowed to build the total-IR (TIR) and FUV 
luminosity functions and densities from $z=0$ to $z=1$ 
\citep[][]{arnouts05,schim,lefloch05,perez05}.
Connecting what is seen in FIR and FUV (rest frame) from low- to high-$z$ is 
crucial to understand above issues, but a new challenge at the same time. 
In this work, we try to examine which population is responsible for
the different evolutions of FUV and FIR galaxies through univariate 
and bivariate luminosity functions (LFs).
At FUV, we made use of {\sl GALEX} datasets, and at FIR, we used {\sl IRAS}
data for the local sample and {\sl Spitzer} data for higher-$z$ ones.
We use the cosmological  parameters 
$H_0 = 72$ km s$^{-1}$ Mpc$^{-1}$, $\Omega_{\rm M} = 0.3$ and 
$\Omega_{\Lambda} = 0.7$.

\begin{figure}[!h]
\centering\includegraphics[clip,width=5.0cm]{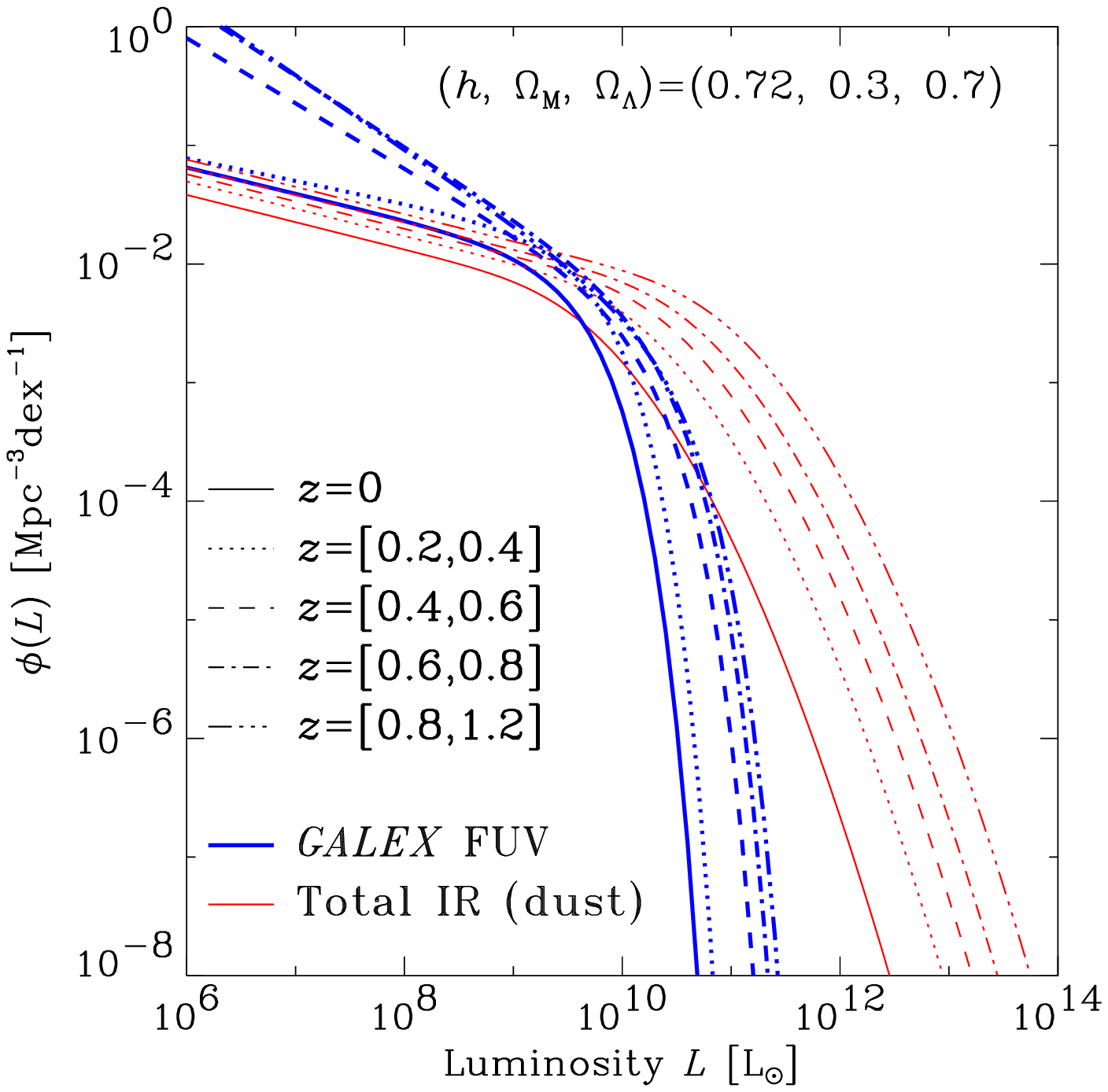}
\centering\includegraphics[clip,width=6.0cm]{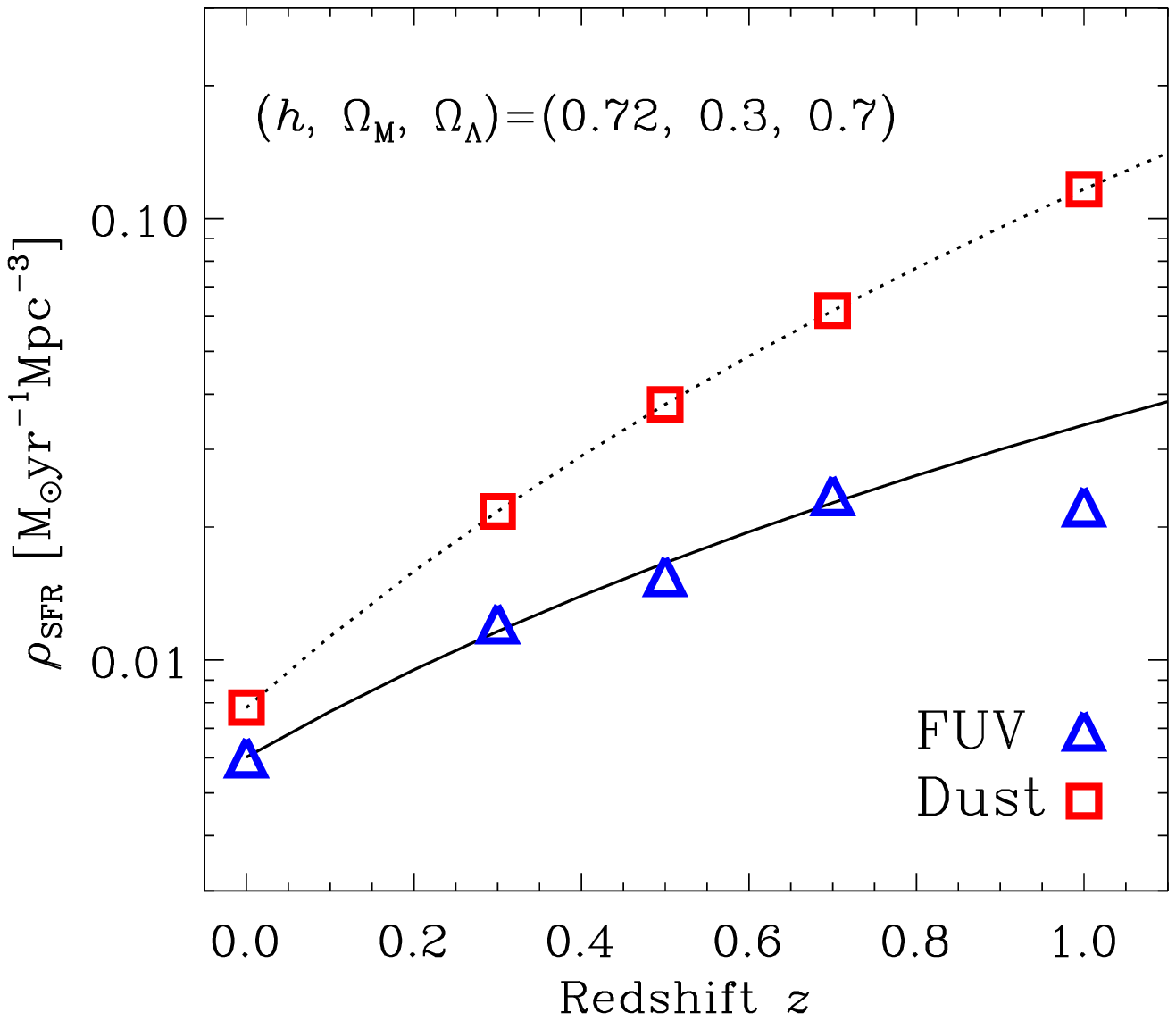}
\caption{Left: The evolution of the FUV and total-IR (dust) luminosity 
functions obtained from {\sl GALEX} and {\sl IRAS/Spitzer}. Right: 
The evolution of the star formation rate density directly obtained from
FUV and total IR luminosity densities.}\label{takeuchi:lf}
\end{figure}

\section{Univariate Analysis}

The evolution of the FUV and dust LFs are shown in Fig.~\ref{takeuchi:lf}.
Local LFs at FUV and FIR are taken from \citet{wyder05} and \citet{takeuchi03},
respectively.
First, it is worth mentioning the (well-known) difference of the local 
LF shape of FUV and dust: for the dust LF, bright galaxies 
($L \ga 10^{10}\;L_\odot$) are much more numerous than those in the FUV.
This leads to the difference in the main population contributing to the total
emitted energy.
In the FUV, the main contributor is $L_*$ galaxies, with fainter galaxies
emitting a non-negligible fraction of energy at $z>0.5$.
In contrast, the effect of the evolution appears in the bright end
for the dust LF.
The contribution from the most luminous galaxies increases with redshift.

By integrating the LFs over luminosities, we obtain luminosity densities 
at FUV and FIR.
Both luminosity densities show a significant evolutionary trend, but the dust 
luminosity density evolves much faster than that of the FUV.
Consequently, the ratio $\rho_{\rm dust}/\rho_{\rm FUV}$ increases toward
higher-$z$, from $\sim 4$ (local) to $\sim 15$ ($z\simeq 1$), 
i.e., the dust luminosity dominates the universe at $z\sim 1$
\citep{takeuchi05a}.

We interpret the data in terms of SFR.
Assuming a constant SFR over $10^8$~yr, and Salpeter initial mass function 
(IMF) \citep[][mass range: $0.1\mbox{--}100\;M_\odot$]{salpeter55}, 
Starburst99 \citep{leitherer99} gives the relation between the SFR and 
$L(\mbox{FUV})\equiv \nu L_\nu$ at FUV (1530~\AA),
\begin{eqnarray}\label{eq:conv_SFR_uv}
  \log L(\mbox{FUV}) = 9.51 + \log \mbox{SFR} \;.
\end{eqnarray}
For the IR, to transform the dust emission to the SFR, we assume
that all the stellar light is absorbed by dust.
Then, we obtain the following formula under the same assumption for both 
the SFR history and the IMF as those of the FUV, 
\begin{eqnarray}\label{eq:conv_SFR_ir}
  \log L(\mbox{dust}) = 9.75 + \log \mbox{SFR} \;.
\end{eqnarray}
However, a significant fraction of the dust emission is due to the heating of 
grains by old stars which is not directly related to the recent SFR.
\citet{hirashita03} found that about 30~\% of the dust heating in the nearby
galaxies comes from stars older than $10^8$~yr.
Adopting this correction, we obtained the evolution of the star formation 
rate densities from FUV and dust ($\rho_{\rm SFR}({\rm FUV})$ and 
$\rho_{\rm SFR}({\rm dust})$) which are presented in 
Fig.~\ref{takeuchi:lf}.
We clearly see that the fraction of hidden SFR increases with redshifts,
and at $z \sim 1$ more than $80\; \%$ of the cosmic SFR is hidden
by dust.

\section{Bivariate Analysis: Star-formation LF in the Local Universe}

We could see that most of the star-forming activity 
in the Universe is hidden by dust at $z \sim 1$.
Then, a natural question arises: which population of galaxies
is due to this evolution?
{}To explore this issue further, we made bivariate LF analysis at
FUV and FIR.
First we have carefully constructed FUV and FIR-selected samples
based on {\sl GALEX} and {\sl IRAS} data.
Using these datasets, we estimated the distribution of the total
luminosity from young stars (star-formation luminosity).
Again following \citet{hirashita03}, we account for the dust heating by 
stars older than ~100 Myr.
Then the star-formation luminosity is 
expressed as $L_{\rm SF} = L_{\rm FUV} +(1-\eta) L_{\rm TIR}$ where 
$\eta$ is the fraction of the TIR emission by old stars.
We adopt $\eta =0.3$ \citep[][]{hirashita03,iglesias06}. 
Therefore $L_{\rm SF}$ can be written as 
$L_{\rm SF} = L_{\rm FUV} +0.7 L_{\rm TIR}$.  
The star formation luminosity function from young stars is calculated for
each sample using the $1/V_{\rm max}$ weighting method. 
For details of the upper limit treatment, see \citet{buat06}.
We found that the $60\;\mu$m luminosity is a robust tracer of the 
luminosity of young stars, while the FUV flux alone (without any correction) 
misses a large part of the total emission of the FUV selected galaxies. 
This trend is related to the relation found between the luminosity 
(or star formation rate) of the galaxies and their dust attenuation 
\citep[e.g.,][]{buat98}.
Both luminosity functions are consistent for 
intermediate luminosities: in the nearby universe these galaxies are 
detected equally well in FIR and in FUV. 
For $L_{\rm SF} \geq 5 \times 10^{10} L_\odot$, the star-formation LF 
issued from the FIR selection is higher than that from the 
FUV one and the discrepancy increases with the luminosity: we miss 
intrinsically bright galaxies which appear much fainter in FUV. 
Deeper wide-area surveys like those of {\sl AKARI} are waited for further
analysis.
Details of the sample construction and analysis are thoroughly 
explained by \citet{buat06}.
Then, how is the situation at higher redshifts?
We tried to explore it by using a Lyman-break galaxy sample at
$z \sim 1$ in the next section.

\section{Bivariate Analysis: Lyman-break galaxies at $z=1$}

\begin{figure}[!h]
\centering\includegraphics[width=5.0cm]{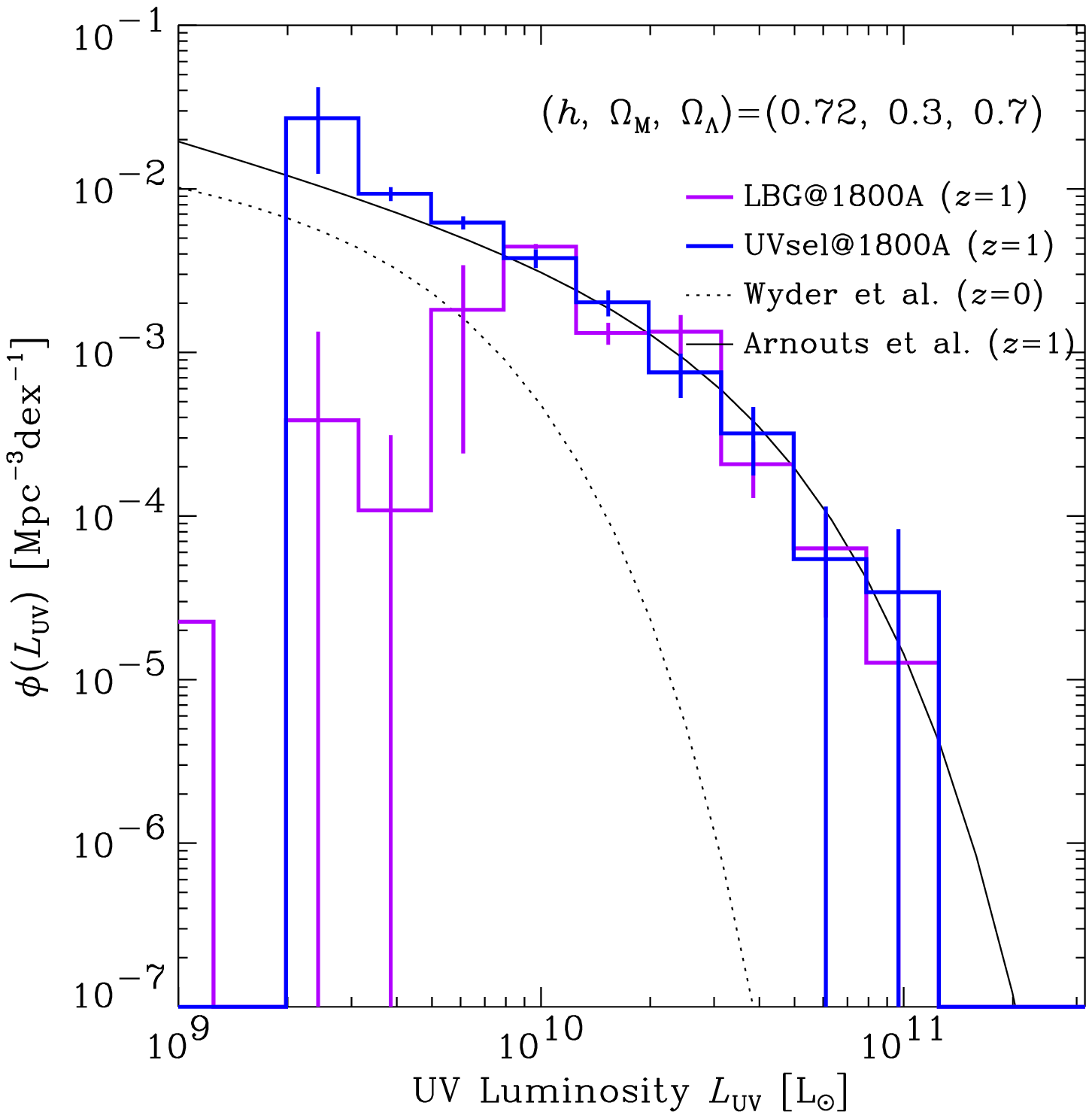}
\centering\includegraphics[width=5.0cm]{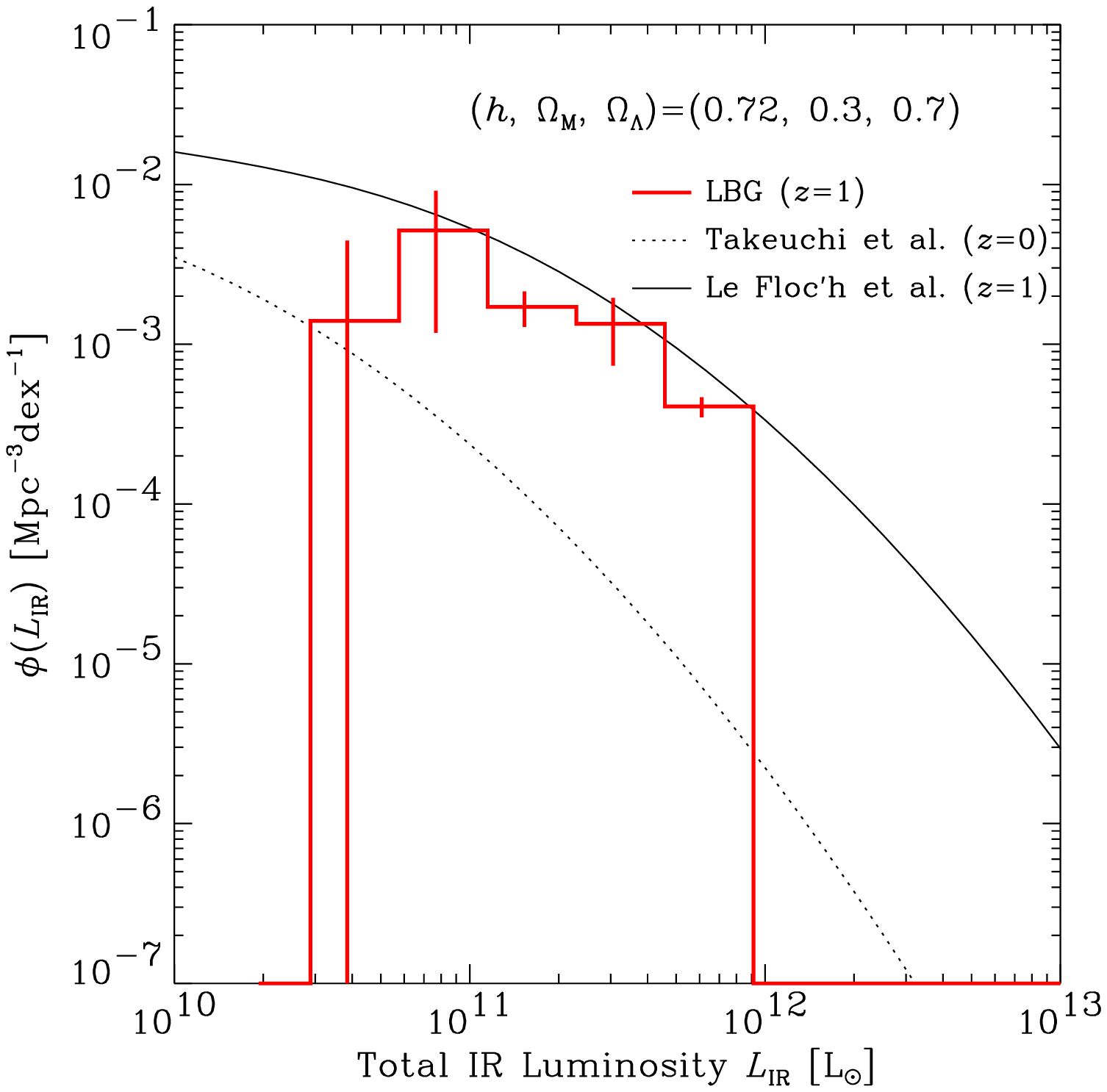}
\caption{The LFs of LBG sample at $z=1$. Left: the total FUV LFs of 
FUV-selected galaxies and LBGs. Right: the dust LF of  LBGs.
Error bars represent 2$\sigma$ (95-\% CL).}\label{takeuchi:lbglf}
\end{figure}

Lyman-break galaxies (LBGs) at $z \sim 1$ are selected by making use of 
{\sl GALEX} FUV and NUV bands \citep{burgarella06}.
It is a suitable method to pick up actively star-forming galaxies at 
these redshifts.
{}To see this, we first estimated the LF of LBGs and compared it with the
total LF of FUV-selected galaxies at $z \sim 1$.
We have used the LBG sample in the Chandra Deep Field South prepared by 
\citet{burgarella06}.
For the LF estimation, we applied $1/V_{\rm max}$ and $C^-$-methods 
according to the recipes of \citet{takeuchi00b} and \citet{takeuchi00a}.
The LFs are shown in Fig.~\ref{takeuchi:lbglf}.
We see that both agree very well at the brighter regime, though LBGs are
much fewer than purely FUV-selected ones at the faint end.

Then, how much is the contribution of the LBGs to the total dust emission?
We can address this issue by estimating the dust LF of LBGs.
For this, we used {\sl Spitzer} MIPS $24\;\mu$m flux to estimate the 
total dust luminosity.
Since at $z \sim 1$ it corresponds to $12\;\mu$m, we made use of the
conversion formula provided by \citet{takeuchi05b}.
The validity of this formula is guaranteed by a comparison with
longer wavelength observation at FIR \citep{takeuchi06}.
In this analysis, we applied the Kaplan--Meier estimator to obtain
the dust LF, to utilize the information of upper limit data properly.
The obtained LF is shown in the right panel of Fig.~\ref{takeuchi:lbglf}.
We found that the dust LF of LBGs is a factor of
2--3 lower than that of the total dust LF of IR-selected galaxies
at the same redshifts \citep{lefloch05}.
This suggests that the evolution of visible galaxies is not
strong enough to explain the drastic evolution of the FIR LF, 
and a FIR-luminous, rapidly diminishing population of galaxies is 
required.

\acknowledgements
We thank the GALEX team.


\begin{thebibliography}{}
\bibitem[Arnouts et al.(2005)]{arnouts05}
 Arnouts, S., et al.\ 2005, \apjl, 619, L43 
\bibitem[Buat \& Xu(1996)]{buat96}
 Buat, V., \& Xu, C.\ 1996, \aap, 306, 61 
\bibitem[Buat \& Burgarella(1998)]{buat98}
 Buat, V., \& Burgarella, D.\ 1998, \aap, 334, 772 
\bibitem[Buat et al.(2005)]{buat05}
 Buat, V., et al.\ 2005, \apjl, 619, L51 
\bibitem[Buat et al.\ (2006)]{buat06}
 Buat, V., et al.\ 2006, \apjs, in press (astro-ph/0609738)
\bibitem[Burgarella et al.(2006)]{burgarella06}
 Burgarella, D., et al.\ 2006, \aap, 450, 69 
\bibitem[Hirashita et al.(2003)]{hirashita03}
 Hirashita, H., Buat, V., \& Inoue, A.~K.\ 2003, \aap, 410, 83 
\bibitem[Iglesias-P{\'a}ramo et al.(2006)]{iglesias06} 
 Iglesias-P{\'a}ramo, J., et al.\ 2006, \apjs, 164, 38 
\bibitem[Le Floc'h et al.(2005)]{lefloch05}
 Le Floc'h, E., et al.\ 2005, \apj, 632, 169 
\bibitem[Leitherer et al.(1999)]{leitherer99}
 Leitherer, C., et al.\ 1999, \apjs, 123, 3 
\bibitem[Schiminovich et al.(2005)]{schim}
 Schiminovich, D., et al.\ 2005, \apjl, 619, L47 
\bibitem[P{\'e}rez-Gonz{\'a}lez et al.(2005)]{perez05} 
 P{\'e}rez-Gonz{\'a}lez, P.~G., et al.\ 2005, \apj, 630, 82 
\bibitem[Salpeter(1955)]{salpeter55}
 Salpeter, E.~E.\ 1955, \apj, 121, 161 
\bibitem[Takeuchi(2000)]{takeuchi00a}
 Takeuchi, T.\ T.\ 2000, \apss, 271, 213
\bibitem[Takeuchi, Yoshikawa, \& Ishii(2000)]{takeuchi00b}
 Takeuchi, T.\ T., Yoshikawa, K., \& Ishii, T.\ T.\ 2000, \apjs, 129, 1
\bibitem[Takeuchi et al.\ (2003)]{takeuchi03}
 Takeuchi, T.\ T., Yoshikawa, K., \& Ishii, T.\ T.\ 2003, \apj, 587, L89
\bibitem[Takeuchi et al.\ (2005a)]{takeuchi05a}
 Takeuchi, T.\ T., Buat, V., \& Burgarella, D.\ 2005a, \aap, 440, L17 
\bibitem[Takeuchi et al.(2005b)]{takeuchi05b}
 Takeuchi, T.\ T., et al.\ 2005b, A\&A, 423, 432 
\bibitem[Takeuchi et al.(2006)]{takeuchi06}
 Takeuchi, T.~T., et al.\ 2006, \aap, 448, 525 
\bibitem[Wyder et al.\ (2005)]{wyder05}
 Wyder, T., et al. 2005, \apjs, 619, L15
\end{thebibliography}
\end{document}